Perspective

# Lessons to learn for better safeguarding of genetic resources during tree pandemics: the case of ash dieback in Europe

Jan Peter George[1], Mari Rusanen[1], Egbert Beuker[2], Leena Yrjänä[1], Volkmar Timmermann[3], Nenad Potocic[4], Sakari Välimäki[2] & Heino Konrad[5]

[1] Production Systems, Natural Resources Institute (Luke), Latokartanonkaari 9, 00790 Helsinki, Finland

[2] Production Systems, Natural Resources Institute (Luke), Vipusenkuja 5, 57200 Savonlinna, Finland

[3] Division of Biotechnology and Plant Health, Norwegian Institute of Bioeconomy Research, 1431, Ås, Norway

[4] Division for Forest Ecology, Croatian Forest Research Institute, Cvjetno naselje 41, 10450, Jastrebarsko, Croatia

[5] Department of Forest Biodiversity and Nature Conservation, Austrian Research and Training Centre for Forests (BFW), Seckendorff-Gudent-Weg 8, 1131 Vienna, Austria

Corresponding author:

Jan Peter George - Production Systems, Natural Resources Institute (Luke), Latokartanonkaari 9, 00790 Helsinki, Finland

Email: jan-peter.george@luke.fi

## Abstract

Ash dieback (ADB) is threatening populations of European ash (*Fraxinus excelsior* & *F. angustifolia*) for more than three decades. Although much knowledge has been gathered in the recent past, practical conservation measures have been mostly implemented at local scale. Since range contraction in both ash species will be exacerbated in the near future by westward expansion of the emerald ash borer and climate change, systematic conservation frameworks need to be developed to avoid long-term population-genetic consequences and depletion of genomic diversity. In this article, we address the advantages and obstacles of conservation approaches aiming to conserve genetic diversity *in-situ* or *ex-situ* during tree pandemics. We are reviewing 47 studies which were published on ash dieback to unravel three important dimensions of ongoing conservation approaches or perceived conservation problems: i) conservation philosophy (i.e. natural selection, resistance breeding or genetic conservation), ii) the spatial scale (ecosystem, country, continent), and iii) the integration of genetic safety margins in conservation planning. Although nearly equal proportions of the reviewed studies mention breeding or active conservation as possible long-term solutions, only 17% consider that additional threats exist which may further reduce genetic diversity in both ash species. We also identify and discuss several knowledge gaps and limitations which may have limited the initiation of conservation projects at national and international level so far. Finally, we demonstrate that there is not much time left for filling these gaps, because European-wide forest health monitoring data indicates a significant decline of ash populations in the last 5 years.

### 1) Introduction

Ash species such as European ash (*Fraxinus excelsior*) and narrow-leafed ash (*F. angustifolia)* are keystone forest tree species characterized by high ecological amplitude, significant economic importance, and vital relevancy for biodiversity. European ash and narrow-leafed ash occur as central functional elements in European riparian forest ecosystems and mountain forests, where they are host species for several hundreds of other taxa (Mitchell et al, 2014). However, both tree species are currently under threat by the invasive pathogen *Hymenoscyphus fraxineus*, the causal agent of ash dieback (ADB), which was first observed in Poland in 1992 and has since spread throughout most of the distribution range of European ash. Lately, it has also reached the southern range margins of both ash species, respectively (Gil et al. 2017; Stroheker et al. 2021; Migliorini et al. 2022). The velocity of spread of ADB is not only remarkable, but also alarming given its relatively young invasion history and small initial population size (McMullan et al. 2019). This has caused that *Fraxinus excelsior*, for example*,* has lately been characterized as "near threatened" at continental scale under the last assessment of The International Union for Conservation of Nature (IUCN) (Khela & Oldfield, 2018) and even as "highly endangered" in national assessments (e.g. Solstad et al. 2021). In addition to this immediate decline which is progressing from the North-East to the South-West of Europe (George et al. 2022), both species are already facing (Rodriguez-Gonzales et al. 2021) or will face (Varol et al. 2021) range contraction at southern range margins because of drier growing conditions. Depletion of genetic diversity because of anthropogenic habitat fragmentation and novel diseases has been recognized as a global problem under the post-2020 global biodiversity framework (Hoban et al. 2023) and specifically implemented in target 4 under the Convention on Biological Diversity: halt species extinction, protect genetic diversity, and manage human-wildlife conflicts. The decision adopted by the conference parties implicitly says: *,,Ensure urgent management actions to halt human induced extinction of known threatened species and for the recovery and conservation of species, in particular threatened species, to significantly reduce extinction risk, as well as to maintain and restore the genetic diversity within and between populations of native, wild and domesticated species to maintain their adaptive potential, including through in situ and ex situ conservation and sustainable management practices,..."* (CBD, 2022). Although there are other species in Europe with a currently higher degree of threat for extinction compared to these two ash species or, which are in general more vulnerable against habitat loss according to the IUCN red list (e.g. *Abies nebrodensis*), there are three key arguments which make conservation efforts in European ash and narrow-leafed ash imperative: i) the velocity of population decline in ash due of ADB is of several magnitudes higher compared to any other tree species with comparable range size in Europe and ii) both species are likely to be potential hosts for the devastating emerald ash borer (EAB), which is spreading westward through Russia and is just about to enter the European Union most likely at the north-eastern boarder where it has been observed lately (Musolin et al. 2021). The third argument is that European ash and narrow-leafed ash, like many other tree species, will face long-term adaptive challenges due to changing climate conditions until the end of this century and these challenges will be most pronounced at range margins: Future species distribution models based on climate envelopes, but also dispersal constraints set by natural migration velocity and land use scenarios (Mauri et al. 2022) predict that until the end of the century both species at the southern range margins will be faced with population contraction because of unsuitable climate conditions (**Supplementary Material S1**).

The massive demographic costs caused by the rapid population decline in ash could even speed up this process several decades as small populations often experience higher drift effects, inbreeding, and limited adaptation potential compared to species with large population sizes (e.g. Lopez et al. 2009; Hoffmann et al. 2017). No matter how bleak these future scenarios may sound, they call for approaches capable of accommodating several dimensions of genetic conservation and rescue which relate to temporal scales (i.e short-term, long-term), spatial scales (ecosystem, country, continent), and also genomic/phenomic scales (e.g. neutral genetic diversity vs. adaptive diversity). These three dimensions are usually reflected in any conservation program aiming to safeguard genetic resources in endangered or threatened species, but they require different planning horizons, measurable success indicators, and evaluation schemes (Figure 1). While temporal scales are usually related to natural turnover rates of the species (e.g. average generation time, time required for a breeding cycle or longevity of seed), spatial scale relates to the space over which sampling of genetic diversity can be realized (e.g. a single stand, a country, or continent-wide). Lastly, the genomic/phenomic scale determines which fraction of the genome-wide or phenome-wide diversity can be captured (e.g. common alleles, rare alleles, field-resistant phenotypes vs. average phenotypes). The genomic/phenomic scale is important in the context of ash because the moderate heritability of resistance against ADB (Enderle et al. 2019 and references therein) and the fact that resistance has a highly polygenic genetic basis (Stocks et al. 2019) will make it necessary to consider genetic safety margins in conservation programs which have a strong focus on resilience breeding. Genetic safety margins will ensure that the species gene pool has enough standing and adaptive genetic variation left for future adaptation processes (e.g. climate) or will be able to withstand future genetic bottlenecks (e.g. impact of the emerald ash borer). What is more, it can be assumed that *Hymenoscyphus fraxineus* has the potential to overcome any genetically determined resistance due to its much shorter generation time and by invasion of new mutants or genotypes (Landolt et al. 2016). In this perspective article, we first report on conservation-related published studies either in European ash or narrow-leafed ash. Specifically, we wanted to know i) whether natural selection, breeding or conservation of genomic resources is the favoured/anticipated long-term solution for the ADB crisis, ii) if there is a bias related to the spatial scale at which the two conservation schemes are applied at, and iii) how many of the proposed conservation solutions take genetic safety margins to other threats into account. Thereafter we discuss pros and cons of different genetic conservation types and their suitability for handling the current ash diversity crisis from various ecological and evolutionary viewpoints. We also unravel research gaps related to genetic conservation, but also point to the fact that a massive amount of publicly available resources has been acquired in recent years for ash which is still underutilized.

2) Literature review on ash dieback from 2011-2023

We queried the Web of Science[TM] search engine with terms "ash dieback" and "conservation" as well as "ash dieback" and "breeding" together with the scientific names for the two ash species in question and got a total of 87 peer-reviewed studies published between 2011 and 2023 (**Supplementary Material S2**). Based on their abstracts, studies were sorted according to the perceived long-term solution for the ADB crisis (natural selection, breeding, conservation, both conservation and breeding), the spatial scale of the study (ecosystem, country, continent), and whether or not other ecological threats such as future invasion of emerald ash borer or climate adaptation were at least theoretically considered as putative causes for further range loss (hereafter referred to as "genetic safety margins"). 47 studies could be assigned to the above defined categories, while the remaining studies did not focus on conservation at all, but were providing technical knowledge such as extraction protocols, genetic markers, etc. Natural selection accounted for the least frequently perceived solution (7 studies) and was associated with small spatial scales (4 at ecosystem and 3 at country level). Breeding and conservation, on the other hand, were mentioned in a larger number of studies (16 and 20, respectively) but differed significantly in the spatial scale of interest: breeding was the preferred solution at country level, whereas conservation is perceived as a solution at ecosystem and country level (Figure 2). Most remarkably, less than 20% of the analysed studies considered the importance of future threats (e.g. emerald ash borer, loss of range due to climate) as additional threats for genetic diversity in both ash species. Range-wide conservation and rescue of genomic resources in ash also seems to be an unrecognized issue as it was mentioned and addressed only in 10 out of 47 studies.

3) Critical review of in situ and ex situ frameworks for conserving genetic resources during tree pandemics

The conservation of genetic resources is generally realized in *in situ* or *ex situ* conservation frameworks. While *in situ* frameworks are always dynamic, because they allow populations to adapt to gradually changing environments by means of generation turnover, *ex situ* frameworks can be both dynamic (e.g. genetic rescue through translocation of populations) or static (e.g. by seed banking). However, most of the frameworks currently in place consider that environmental changes are gradual rather than abrupt as in the case of a tree pandemic. In fact, high demographic costs in the case of ash dieback can violate some of the assumptions behind gradual population adaptation (e.g. Buerger & Lynch 1995) and therefore makes it necessary to review the existing approaches.

Dynamic in situ is the preferred framework for genetic conservation of adaptive diversity in Europe for many of the existing tree species (e.g. Koskela et al. 2013; Lefevre et al. 2013). A network of 146 ash conservation units currently exists at European level (128 for *F. excelsior* and 18 for *F. angustifolia*) covering a substantial part of the European range (Euforgen 2024) and the main objectives of this network are to maintain neutral and adaptive diversity needed for adaptation to climate (DeVries et al. 2013). Since the most important indicator to reach this objective is the number of reproducing ash trees in the units which should be in the range of 50-500, it is clear that ash dieback has the potential to counteract in situ conservation efforts because of the excessive mortality rates in natural ash populations which are observed across all spatial scales (e.g. George et al. 2022; Coker et al. 2019). Rapid reduction in population size almost ever leads to temporal fluctuation in genetic variance and simulation studies have shown that these random fluctuations in genetic variance have the potential to drastically speed-up population extinction during adaptation processes (Buerger & Lynch, 1995). The only possible way to avoid or at least to postpone extinction would be to manage dynamic in situ units to ensure that as many trees as possible contribute equally to natural regeneration within the unit. This means that either the inbreeding effective population size or the variance effective population size will be the most important indicators for conservation success.

Dynamic in situ conservation takes essentially place at ecosystem or metapopulation level (note that the word "network" above does not mean that the in situ units are genetically connected at continental scale). It is therefore the only conservation framework which has the potential to co-conserve other taxa which are directly or indirectly associated with ash (Hultberg et al. 2020; Mitchell et al. 2014). Figure 3 gives an overview of the advantages and challenges for dynamic in situ conservation in case of a tree pandemic: While the strengths of dynamic in situ lies in its ability to co-conserve other taxa and maintain ecosystem functions (e.g. in riparian ecosystems), its weaknesses are the high conservation and monitoring effort which is necessary to achieve high effective population sizes in case of biotic outbreaks and its vulnerability to subsequent biotic damages such as the emerald ash borer.

Dynamic ex situ conservation is usually applied to populations for which genetic rescue is needed (e.g. assisted migration or assisted colonization), but also when populations are formed by too few individuals in the wild. In the latter case so-called synthetic populations can be created in form of conservation stands or conservation orchards to enhance gene flow and increase the effective number of breeders. Dynamic ex situ conservation units could be the logical consequence from any ash breeding program aiming to produce forest reproductive material with higher resilience against ash dieback (e.g. Liziniewicz et al. 2022), because it would allow to assemble field-resistant mother trees on a relatively small area in order to increase cross-pollination and panmixia. The strength of this approach clearly lies in the ability to control effective population size without much monitoring effort by making sure that seed is only harvested when a minimum number of trees is flowering or bearing seed. The weaknesses are two-fold: first, it needs to be ensured that the produced seed is regularly employed elsewhere in restoration projects and that survival among planted families in natural habitats is more or less even. Outbreeding depression and maladaptation can be major issues when field-resistant clones or families are sampled over wide environmental gradients or when the produced seed is employed too far away from its origin during restoration. The second problem of this approach has to do with the selection of resistant or tolerant clones in the field: Since the general assumption is that the heritability in resistance against ash dieback is rather moderate (0.1-0.65; Evans 2019), a relatively strong selection will be necessary when trying to improve this trait by means of classical breeding (Falconer and MacKay 1996). This could seriously narrow-down and even deplete important genetic variation needed for subsequent selection and adaptation processes such as resistance against EAB or new *Hymenoscyphus fraxineus* mutants. Dynamic ex-situ conservation will most likely take place at country-level, where seed deployment zones and national regulations will determine the legal frame for use and deployment of forest reproductive material. However, dynamic ex situ ash populations could even be established without selection for breeding by translocating trees to regions which are not or are less impacted by the disease (e.g. drier sites or islands). In such a case a dynamic ex situ unit could produce seed with higher-than-average genetic diversity and could be used for restoration experiments or simply as back-up in long-term seed repositories.

The latter point leads to static conservation approaches which are -by definition- always realized in *ex situ* collections either by seed or any other kind of germplasm storage. Many conservationists would probably claim that static ex situ is currently the most effective way of conserving and safeguarding ash genomic resources at continental scale while keeping as many evolutionary options open as possible. Static ex situ approaches can be powerful both before and during tree pandemics because they allow to sample genetic variation over wide geographic scales with relatively low time effort (Hoban et al. 2018). European ash seed are orthodox and can be stored over long time (Chmielarz 2009) in forest seed banks. The big strength of static ex situ collections is the flexibility to design the sampling process: depending on the defined conservation goal and environmental frame conditions the conservationist can calculate the most efficient number of populations or mother trees to be sampled (Hoban 2019; Hoban et al. 2018). However, the sampling effort may vary substantially depending on the goal (see next section). Static ex situ collections are able to capture almost unlimited amounts of species genomic diversity (e.g. rare alleles, adaptive diversity at range margins) and will be most effective in creating safety margins for coming threats. However, compared with in situ approaches static ex situ collections have no opportunity to leverage the force of natural selection for adaptation to changing environments. They also require critical infrastructure and knowledge for maintaining and regular testing of viability of the stored germplasm, which is not always at hand. To give an idea of the efforts which would be imaginable at country level, the UK national tree seed project (NTSP) can serve as an example: 2.3 million ash seed have been collected across the UK and are currently stored in the Millenium Seed Bank in Wakehurst (Hoban et al. 2018). This collection accounts for approximately 90% of the expected allelic diversity in ash in the UK.

Although these three approaches are here discussed separately from each other, several combinations between dynamic and static or between *in situ* and *ex situ* approaches are conceivable, but maybe still underutilized. Seed banking, cryopreservation, and the establishment of synthetic *ex situ* populations could be, for example, utilized to complement dynamic in situ approaches. In this way, effective population sizes could be to some degree maintained *in situ* by reforestation with autochthonous plant material originating from ex situ collections. The ecological and economical importance of the ecosystem in question will determine if such a combined approach is a worthy investment, but thoughts should be given to these ideas before giving up any ash stand or ecosystem entirely.

4) Research gaps and underutilized resources

While the conservation workflow illustrated in Figure 1 gives a broad overview of the steps which are necessary to establish a conservation program in face of a tree pandemic, it also raises questions about the availability of information in order to derive the required metrics, particularly for defining measurable success indicators (Figure 1c) and for deriving risk and contingency margins (Figure 1 d). For dynamic in situ conservation, the so-called 50/500 rule was and is still used as a proxy for minimum population viability (Franklin 1980, but see Frankham et al. 2014 for a critical review). In simple words, the rule says that a short-term effective population size of 50 will be sufficient to buffer against the negative effects of inbreeding, while a long-term effective population size of 500 individuals will be able to balance the loss of adaptive genetic variation due to drift effects. However, the conceptual nature of $N_e$ makes it impractical to be used as an operational verifier for practical conservation in cases where populations are already under severe decline or under immediate threat of severe decline as in the case of ADB or EAB (Jamieson & Allendorf 2012). We see therefore a clear need to develop and evaluate novel and applied indicators for monitoring in situ conservation success which are capable to provide verifiers for minimum viable population sizes in cases of rapid population decline. One way to achieve this could be to collaborate more strongly in the future with the scientific community involved in genetic simulations. In fact, simulation studies have been proven useful to model complex evolutionary and ecological scenarios for guiding practical conservation (e.g. Hoban 2019; Hoban & Schlarbaum 2014; Hoban et al. 2012). What is more, genetic simulation frameworks have been improved recently and permit nowadays for modelling multi-species eco-evolutionary dynamics over short and long time scales (Haller & Messer 2023). We strongly believe that practical conservation genomics concerned with the management of biotic outbreaks could substantially benefit from the recent achievements in genetic simulations and that such applications could shed light on the co-evolutionary perspective between ash and its damaging agents, particularly ADB and EAB. Given that excellent and freely available genomic resources exist for both European ash and narrow-leafed ash (Sollars et al. 2017; Kelly et al. 2020), evolutionary consequences of rapid population decline could therefore be reliably estimated from the community-level down to gene level. Research resources such as newly developed markers are already applied in existing field experiments to study the population genetic consequences of ash populations exposed to ADB: Metheringham et al. (2022) found clear evidence of allele frequency changes related to selection within one generation, which demonstrates that micro-evolution in natural ash populations is already happening. We see the value of such *in situ* experiments in two ways: first, they can be used for establishing conservation baseline scenarios in stands which are already facing a long ADB history without any human intervention (e.g. Old growth natural ash stands in eastern Europe such as Białowieża primeval forest) to see how natural selection alone will shape the future ash genome. Second, they can be used as starting values or priors for forward simulations under complex eco-evolutionary scenarios to guide future conservation programs (both *in situ* and *ex situ*). In this regard, it is of utmost importance that European research infrastructure will be further developed to enable sharing of data and knowledge across borders. Some European initiatives have been developed in this context such as the European COST action Fraxback (https://www.cost.eu/actions/FP1103/) and also international training workshops such as the ash genome workshop (https://www.kew.org/read-and-watch/accelerating-ash-breeding-programmes). More such initiatives need to be promoted through European research programs to ensure an efficient exchange between European researchers and to accelerate conservation efforts on ash.

However, at country- and ecosystem level, success indicators and safety margins need to be assessed differently, since these geographic scales often align with different conservation frameworks (Figure 3): while dynamic ex situ and static ex situ conservation will both be feasible at country level (e.g. conservation seed orchards, national seed banks), static ex situ will be the most likely approach at continental scale (e.g. international seed or clone banks), because of its cost-effectiveness. The challenge of defining success indicators and safety margins for conservation at continental scale lies in the fact that range-wide loss of genetic diversity is not only happening because of biotic outbreaks such as ADB, but generally as a consequence of man-made range reduction in almost all forest trees (Exposito-Alonso et al. 2022; Mauri et al. 2022). Therefore, three quantities need to be known to define a success indicator for conservation at continental scale: first, how genetic diversity is distributed in space, second, the spatio-temporal sequence of population decline, and third, how much of the loss in genetic diversity is attributable to biotic outbreaks in comparison to other processes (i.e. the baseline loss). If the first quantity is unknown, it will be impossible to define efficient sampling strategies (e.g. number of individuals necessary from different environmental zones or conservation areas). If the second quantity is unknown, it is difficult to prioritize areas for conservation (e.g. northern range margins are under stronger decline, because ADB arrived there earlier). If the third quantity is unknown, it will not be possible to evaluate conservation success, because the baseline for comparison is unknown or constantly changing. Not knowing the second and third quantity will result in a so-called shifting baseline syndrome (Exposito-Alonso 2023) and will most likely underestimate the estimated loss and consequently efforts necessary to conserve genetic diversity. Loosely speaking, without knowledge about the aforementioned frame conditions, any ex situ conservation effort at larger geographical scale will fail to prove that it has captured enough genetic diversity able to balance the anticipated loss. Figure 4 gives a schematic and simplified overview for the relationship between range loss and loss in genetic diversity and illustrates how baseline loss and loss due to biotic outbreaks may differ. Although it seems like a mammoth task to establish an accurate mathematical relationship between geographic range loss and genetic diversity, similar relationships do exist for a wide array of plant and tree species which may serve as an inspiring template for ash (Exposito-Alonso et al. 2022). Such research will need to take advantage of various other research resources ranging from long-term forest inventory data, remote sensing at different temporal and spatial resolution, and accurate species distribution models and we believe that crosstalk between these research disciplines and genetics is of utmost importance for the future conservation of threatened species. Many of those needed research resources actually exist for European ash and Narrow-leafed ash as we have outlined above and will further show in section 5, but they have not been utilized yet collectively.

Finally, there are still research gaps which relate to practical conservation issues, but which need to be tackled in the near future. As such, cryopreservation has been proven a cost-effective method in trees for ex situ long-term preservation of germplasm (Häggman et al. 2008). Cryopreservation of dormant buds with subsequent micropropagation has successfully been applied in wych elm and white elm in Finland (Välimäki et al. 2021). Similar cryocollection with dormant buds from wych elm (*Ulmus glabra*), white elm (*Ulmus laevis*) and field elm (*Ulmus minor*) collected from nine European countries has been established as duplicates in Germany and in France (Harvengt et al. 2004, Collin et al. 2020). These species are facing a similar fate in Europe as ash since the Dutch Elm Disease has killed a considerable part of the populations since the beginning of the 20[th] Century (Brasier 1991). Cryocollections can function as an alternative or complementary conservation method to seed banking, where a large number of genotypes can be securely stored long-term with minimal space requirements. Dormant buds of ash species can be cryopreserved and regenerated by grafting (Volk et al. 2009) while micropropagation protocols (e.g. Hammat and Ridout 1992; Šedivá et al. 2015) have also been developed.

However, regeneration through micropropagation can be still challenging due to the high likelihood for microbial contaminations in ash (Donnarumma et al. 2010) and attention should be given for improving this technique. One option could also be to use seed embryos as explants for somatic embryogenesis (Capuana et al. 2007, Merkle et al. 2022), which is compatible with cryopreservation (Richins et al. 2024). However, this would not enable the preservation of adult trees, but their progeny. On the other hand, seed embryos are easier to sterilize than dormant buds, and somatic embryogenesis is generally more scalable than other tissue culture methods.

5) A current picture on ash decline from a continental perspective

An important question during biotic outbreaks is how much time is left to put a genetic conservation program into action before critical amounts of genetic diversity are lost. Several countries have reported on regional or national population decline during the last three decades (e.g. Timmermann et al. 2017; Enderle et al. 2018; Cleary et al. 2017) and subsequently various conservation efforts were initiated at national level (see for example the Austrian initiative Ash in distress (https://www.esche-in-not.at) or the British ash conservation initiative (https://www.gov.uk/government/publications/ash-tree-research-strategy-2019/conserving-our-ash-trees-and-mitigating-the-impacts-of-pests-and-diseases-of-ash-a-vision-and-high-level-strategy-for-ash-research)). Figure 5 shows how the decline in European ash and narrow-leafed ash populations has advanced since the arrival of ADB in Europe. The data is based on annual observations within the ICP Forests Level I survey (Schwärzel et al 2022, Eichhorn et al. 2020, George et al. 2022), which systematically assesses tree vitality on a 16x16km grid across Europe providing valuable and detailed field information on crown defoliation and biotic and abiotic damage (Potočić et al. 2021, Timmermann et al. 2023). In total, more than 400 plots exist in which either *Fraxinus excelsior* or *Fraxinus angustifolia* trees were repeatedly surveyed before and after arrival of ADB in Europe. This dataset unravels an interesting pattern of decline: mortality in ash was not significantly different from the baseline mortality in the same plots (baseline was defined as mortality of all other tree species in the same plots) until 2013. In contrast to this, the data suggest a sharp and monotone increase in mortality after 2015 which can be best described as an exponential function which clearly exceeds the baseline mortality level. The maps in Figure 5 also reveal that immediate conservation priority should be given to northern range margins as they are under high decline pressure, while core populations are currently experiencing smaller, but yet significant population decline. The temporal dynamics unraveled by the long-term observation network illustrates that European ash and narrow-leafed ash are possibly trapped in an extinction debt: although mortality appeared to be below the signal-to-noise ratio until 2010, this picture may have been masked by the fact that it normally takes several years for large, dominant trees to be killed by the pathogen. Therefore, the time lag between pathogen arrival and first signals of excess mortality at continental scale could have been used by conservation programs for safeguarding valuable genetic resources in ash. Conservation programs should start before genetic consequences of population isolation occur (Lopez et al. 2009) and based on the ICP Forests Level I monitoring data it seems imperative that these actions should start very soon across the continent, if not already done. A continent-wide monitoring such as ICP Forest Level I can never give the complete picture of decline, because of its relatively coarse spatial resolution. Since European ash is a rather rare species in the North and Narrow-leafed ash is rare throughout its distribution, mortality estimates may not always reveal a fully representative picture, because the probability that ash will occur in the ICP plots is strongly correlated with its abundance or density in the survey area. Nevertheless, the data unveils that mortality patterns are generally in congruence with the spatio-temporal progression of ADB and follow a north-to-south gradient. The pattern seem to be also in line with national assessment schemes in the North, which have recently re-classified European ash as an endangered to highly threatened species (Solstad et al. 2021; Sundberg 2020). Consequently, conservation programs should prioritize preservation of neutral and adaptive genetic variation at northern range margins first.

6) Conclusions

The aim of this article was to open up a perspective for future conservation of genomic resources threatened not only by gradual environmental changes, but also by rapid population decline due to biotic outbreaks, with ash dieback serving as an example. Prospects for European ash and narrow-leafed ash are anything but promising, given the reported expansion velocities for the emerald ash borer in Europe (Musolin et al. 2021) and the fact that some of the most important native American ash species such as *Fraxinus americana, Fraxinus pensylvanica, and Fraxinus caroliniana* have been lately assessed as endangered and critically endangered species, respectively, as a result of EAB (Jerome et al. 2017). There is currently no single solution or best practice for handling these kinds of situations. We were able to identify strengths and weaknesses of the three main conservation approaches (dynamic *in situ*, dynamic *ex situ*, static *ex situ*) and how they align with spatial scales at which conservation is applied. Although natural selection is still perceived as a solution according to the literature we have reviewed, active conservation seems to be the preferred way of handling the ash crisis in the majority of studies. However, a pan-European conservation strategy for ash, is still missing probably because of some of the limitations described in this article. Practically speaking, the start of a pan-European conservation strategy should not wait until all missing theoretical frame conditions have been resolved, but rather use the wealth of acquired resources and knowledge to make important steps forward in this direction. The large heterogeneity in conservation approaches at European level (for example in phenotyping, germplasm banking, selection and breeding, etc.) would allow to synthesize results quickly to inform those countries which are not yet affected by ash dieback, but will be in the near future. It is also desirable to look back in time and try to learn from past initiatives which aimed to safeguard European-wide genetic variation during the devastating Dutch elm disease (e.g. Collin et al. 2004) in order to avoid mistakes that were already made.

Landolt, J., Gross, A., Holdenrieder, O., & Pautasso, M. (2016). Ash dieback due to Hymenoscyphus fraxineus: what can be learnt from evolutionary ecology?. Plant Pathology, 65(7), 1056-1070.

Lefevre, F., Koskela, J., Hubert, J., Kraigher, H., Longauer, R., Olrik, D. C., ... & Zariŋa, I. (2013). Dynamic conservation of forest genetic resources in 33 European countries. Conservation Biology, 27(2), 373-384.

Liziniewicz, M., Tolio, B., & Cleary, M. (2022). Monitoring of long-term tolerance of European ash to Hymenoscyphus fraxineus in clonal seed orchards in Sweden. Forest Pathology, 52(5), e12773.

Lopez, S., Rousset, F., Shaw, F. H., Shaw, R. G., & Ronce, O. (2009). Joint effects of inbreeding and local adaptation on the evolution of genetic load after fragmentation. *Conservation Biology*, *23*(6), 1618-1627.

Mauri, A., Girardello, M., Strona, G., Beck, P. S., Forzieri, G., Caudullo, G., ... & Cescatti, A. (2022). EU-Trees4F, a dataset on the future distribution of European tree species. Scientific Data, 9(1), 37.

McMullan, M., Rafiqi, M., Kaithakottil, G., Clavijo, B. J., Bilham, L., Orton, E., ... & Clark, M. D. (2018). The ash dieback invasion of Europe was founded by two genetically divergent individuals. Nature Ecology & Evolution, 2(6), 1000-1008.

Merkle, S. A., Koch, J. L., Tull, A. R., Dassow, J. E., Carey, D. W., Barnes, B. F., ... & Gandhi, K. J. (2023). Application of somatic embryogenesis for development of emerald ash borer-resistant white ash and green ash varietals. *New Forests*, *54*(4), 697-720.

Migliorini, D., Luchi, N., Nigrone, E., Pecori, F., Pepori, A. L., & Santini, A. (2022). Expansion of Ash Dieback towards the scattered Fraxinus excelsior range of the Italian peninsula. Biological Invasions, 24(5), 1359-1373.

Mitchell, R. J., Beaton, J. K., Bellamy, P. E., Broome, A., Chetcuti, J., Eaton, S., ... & Woodward, S. (2014). Ash dieback in the UK: a review of the ecological and conservation implications and potential management options. Biological conservation, 175, 95-109.
20

**Figure captions:**

**Fig 1:** A schematic workflow of a conservation program during a tree pandemic and how steps align with spatial scale of conservation. Grey boxes give examples for each implementation step. a) Impact assessment can make use of any data resource, presumably available as time series or historical record. [1]ICP forest level I crown defoliation dataset (http://icp-forests.net/), [2] European Environmental Agency (https://www.eea.europa.eu/), [3] Joint Research Centre (https://www.data.jrc.ec.europa.eu/). b) Inventory and review will help to prioritize conservation and will help to synthesize with ongoing conservation programs. c) Conservation objectives need to be aligned with measurable indicators, which will make it possible to measure success or failure of a conservation program. Indicators don't need to be necessarily numeric, but should be scientifically backed. d) A risk & contingency plan will make sure that sufficient safety margins are considered. These safety margins can, for example, relate to other abiotic or biotic threats such as climatic extremes or forthcoming pandemics (e.g. the emerald ash borer). e) Implementation of the conservation strategy in one or more of the available conservation frameworks. f) Evaluation needs to ensure that defined goals have been reached or that adjustments are implemented in case indicators suggest conservation deficits. Boxes in red are essential, but also the most unknown for which a lot of knowledge gaps currently exist.

**Fig. 2:** Results of the literature review. Number of studies and their perceived conservation strategy. Natural selection can be seen as a baseline for which no active conservation is thought to be needed. Safety margins were broadly defined as the perceived necessity to conserve genetic resources more broadly and in face of upcoming threats for populations (e.g. climate, emerald ash borer). ESY: Ecosystem; CNT: Country; CON: Continent

**Fig. 3:** Strengths and weaknesses of the three main conservation types at the three spatial scales (ecosystem, country, continent). Relative conservation effort will decrease with increasing spatial scale and the more static the approach will be. Long-term adaptive capacity will be ensured for dynamic in situ and, under certain circumstances, for dynamic ex situ. Insurance margins can be realized most efficiently under static ex situ, because a nearly unlimited number of specimens can be sampled and stored. Co-conservation of species associated with the declining host will be possible only in-situ.

**Fig. 4:** A schematic relationship between species range loss and loss in genetic diversity as proposed by Exposito-Alonso (2023). Both anthropogenic range loss (black) and loss due to ash dieback (red) will come with a high uncertainty which is illustrated by grey and orange lines, respectively. The main reason is that anthropogenic range loss and range loss due to the tree pandemic are spatially correlated, respectively. For instance, evidence suggests that ash is loosing parts of its natural range in the south due to climactically induced extinction (e.g. Mauri et al. 2022). In contrast, northern range margins experiencing contraction because of ADB (e.g. George et al. 2022), while the pathogen arrived only lately at southern range margins. Although it is still difficult to disentangle and accurately assess range loss through time, a lot of historic and contemporary data sources exists which should be utilized more strongly in the future.

**Fig. 5:** Data from long-term forest health monitoring such as the ICP Forests level I dataset is able to monitor population decline at European level. The maps show how mortality in European ash and narrow-leafed ash has advanced from 1987-2023 in Europe. Size of the circles is proportional to mortality rates. The same data is shown as points as European-wide average together with background mortality as a baseline. The graphics illustrate that despite the fact that ADB has arrived already in the beginning of the 1990s, it took considerable time until population decline has become visible at European scale. Ash mortality points (red) were fitted with an exponential function that explained approx. 40% of the temporal variation in mortality. A total of 407 ICP Forests Level I plots across Europe containing *Fraxinus excelsior* and *Fraxinus angustifolia* were analyzed for producing this graph. Note that despite the high mortality in Norway and Sweden only a very low number of ash trees were present in these plots between 2011 and 2023.

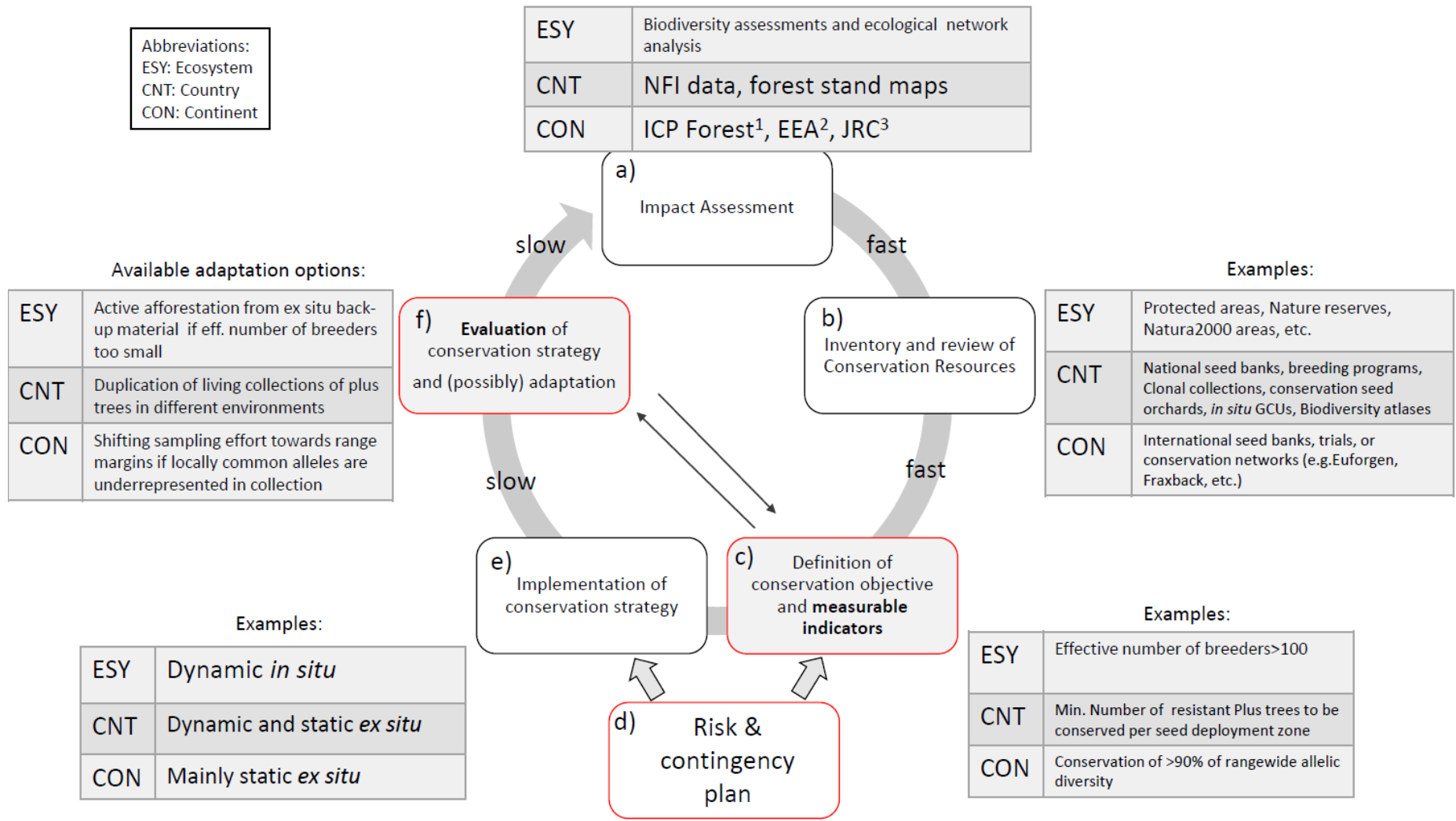

**Fig 1:** A schematic workflow of a conservation program during a tree pandemic and how steps align with spatial scale of conservation. Grey boxes give examples for each implementation step. a) Impact assessment can make use of any data resource, presumably available as time series or historical record.

[1]ICP forest level I crown defoliation dataset (http://icp-forests.net/), [2] European Environmental Agency (https://www.eea.europa.eu/), [3] Joint Research Centre (https://www.data.jrc.ec.europa.eu/). b) Inventory and review will help to prioritize conservation and will help to synthesize with ongoing conservation programs. c) Conservation objectives need to be aligned with measurable indicators, which will make it possible to measure success or failure of a conservation program. Indicators don't need to be necessarily numeric, but should be scientifically backed. d) A risk & contingency plan will make sure that sufficient safety margins are considered. These safety margins can, for example, relate to other abiotic or biotic threats such as climatic extremes or forthcoming pandemics (e.g. the emerald ash borer). e) Implementation of the conservation strategy in one or more of the available conservation frameworks. f) Evaluation needs to ensure that defined goals have been reached or that adjustments are implemented in case indicators suggest conservation deficits. Boxes in red are essential, but also the most unknown for which a lot of knowledge gaps currently exist.

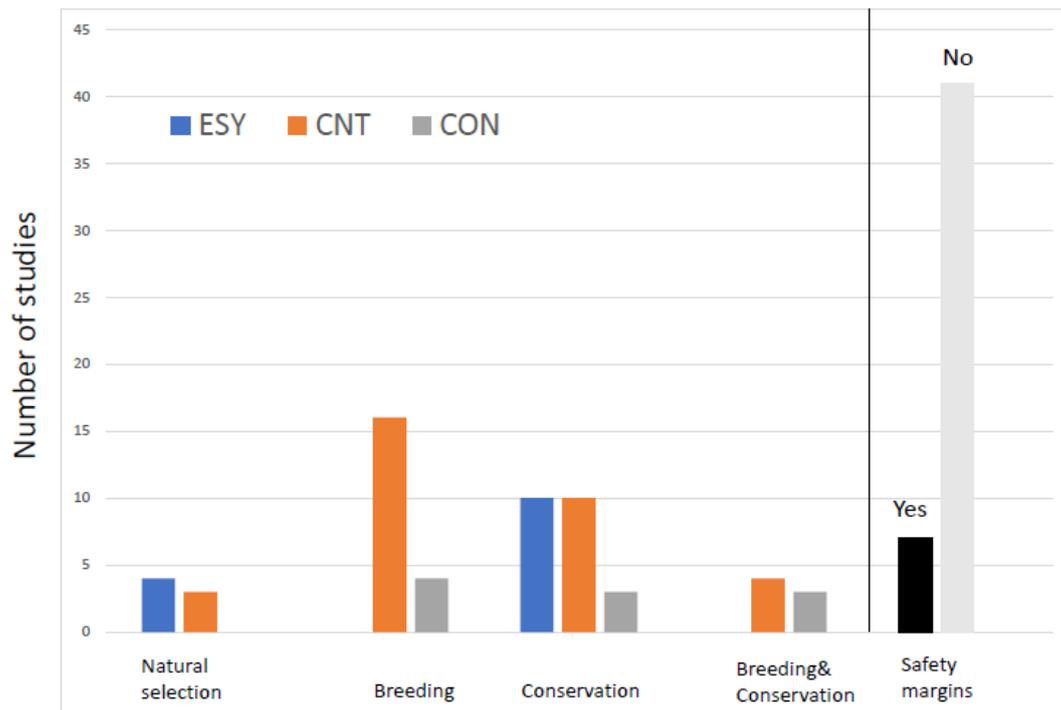

**Fig. 2:** Results of the literature review. Number of studies and their perceived conservation strategy. Natural selection can be seen as a baseline for which no active conservation is thought to be needed. Safety margins were broadly defined as the perceived necessity to conserve genetic resources more broadly and in face of upcoming threats for populations (e.g. climate, emerald ash borer). ESY: Ecosystem; CNT: Country; CON: Continent

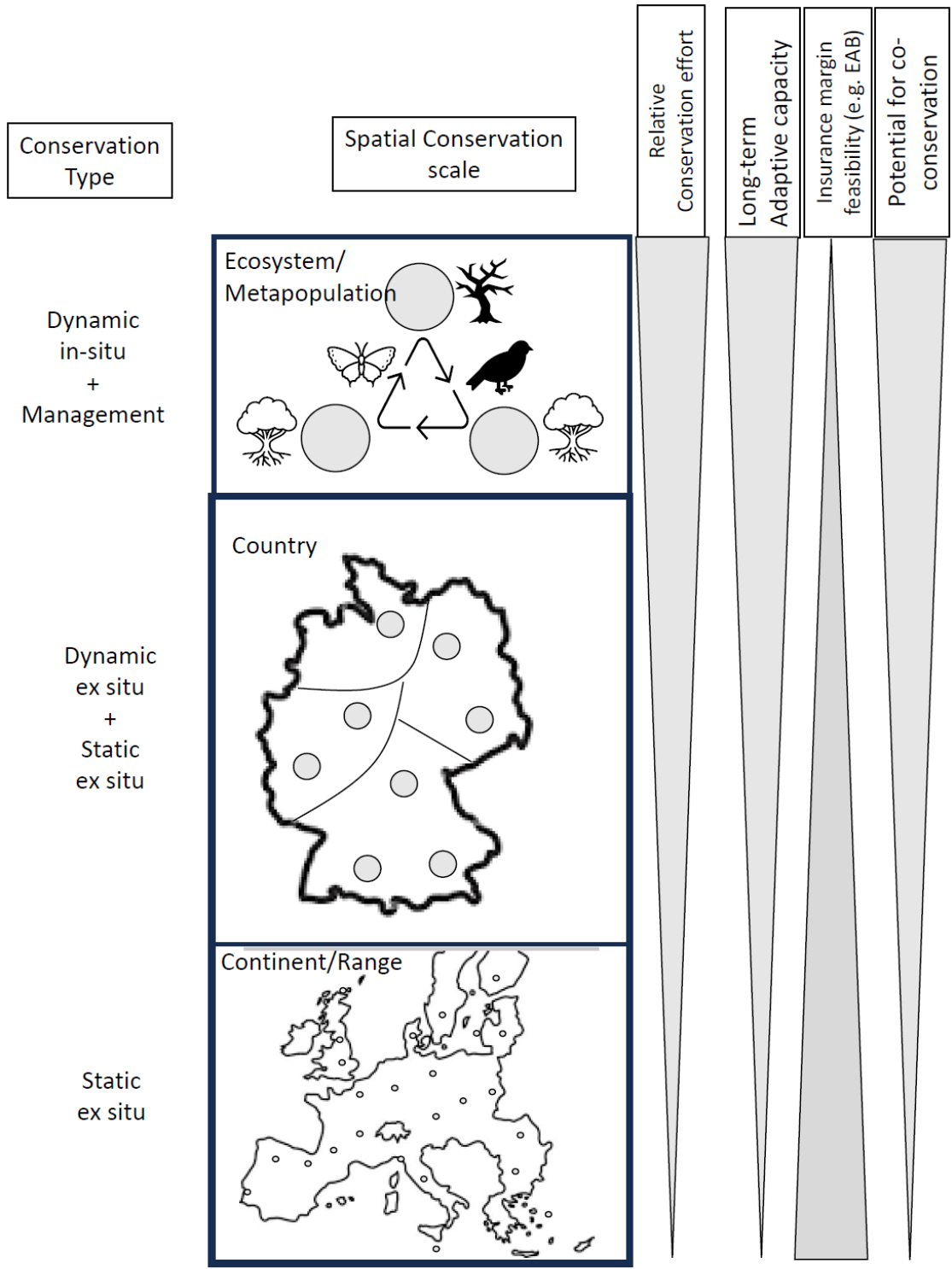

**Fig. 3:** Strengths and weaknesses of the three main conservation types at the three spatial scales (ecosystem, country, continent). Relative conservation effort will decrease with increasing spatial scale and the more static the approach will be. Long-term adaptive capacity will be ensured for dynamic in situ and, under certain circumstances, for dynamic ex situ. Insurance margins can be realized most efficiently under static ex situ, because a nearly unlimited number of specimens can be sampled and stored. Co-conservation of species associated with the declining host will be possible only in-situ.

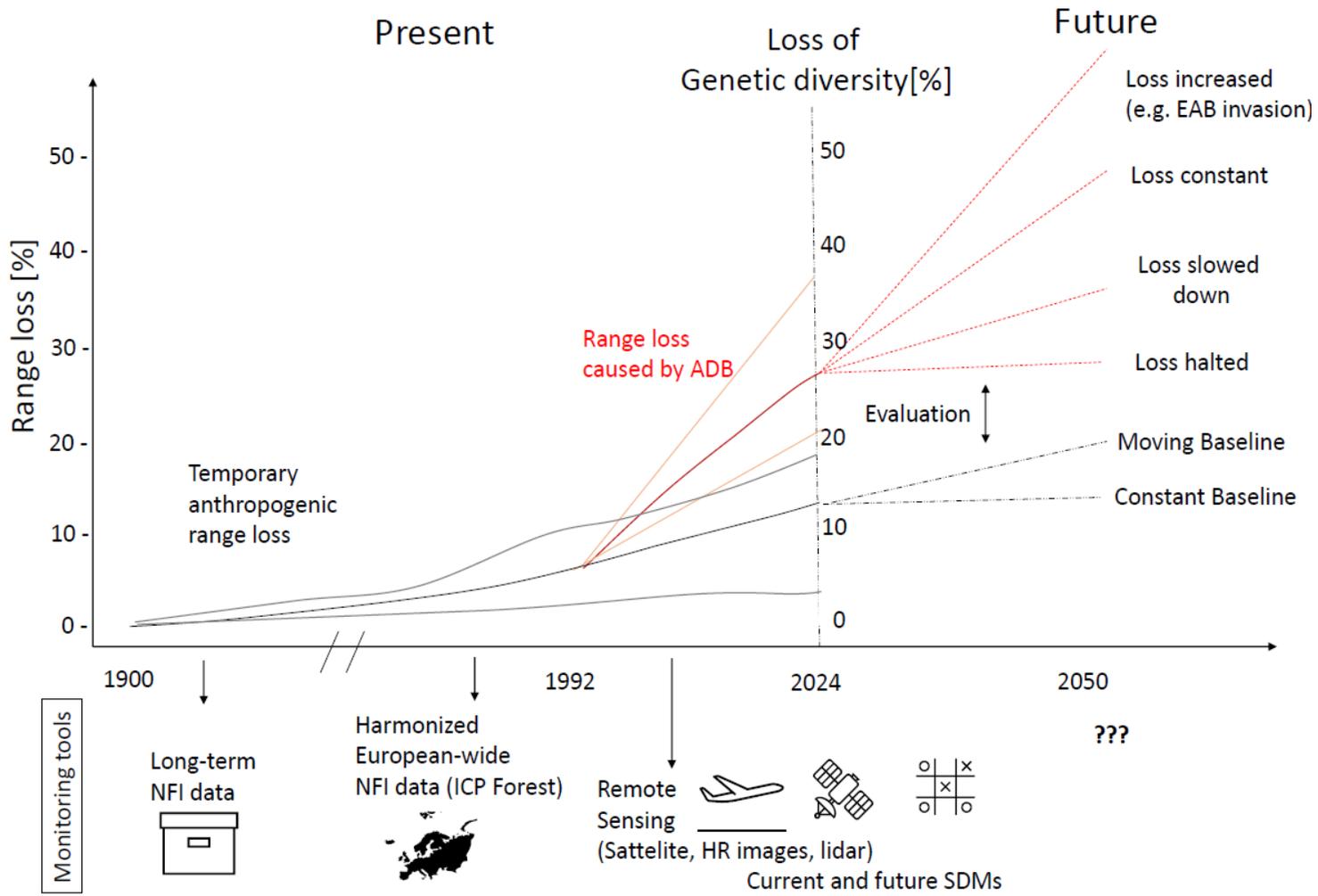

**Fig. 4:** A schematic relationship between species range loss and loss in genetic diversity as proposed by Exposito-Alonso (2023). Both anthropogenic range loss (black) and loss due to ash dieback (red) will come with a high uncertainty which is illustrated by grey and orange lines, respectively. The main reason is that anthropogenic range loss and range loss due to the tree pandemic are spatially correlated, respectively. For instance, evidence suggests that ash is loosing parts of its natural range in the south due to climactically induced extinction (e.g. Mauri et al. 2022). In contrast, northern range margins experiencing contraction because of ADB (e.g. George et al. 2022), while the pathogen arrived only lately at southern range margins. Although it is still difficult to disentangle and accurately assess range loss through time, a lot of historic and contemporary data sources exists which should be utilized more strongly in the future.

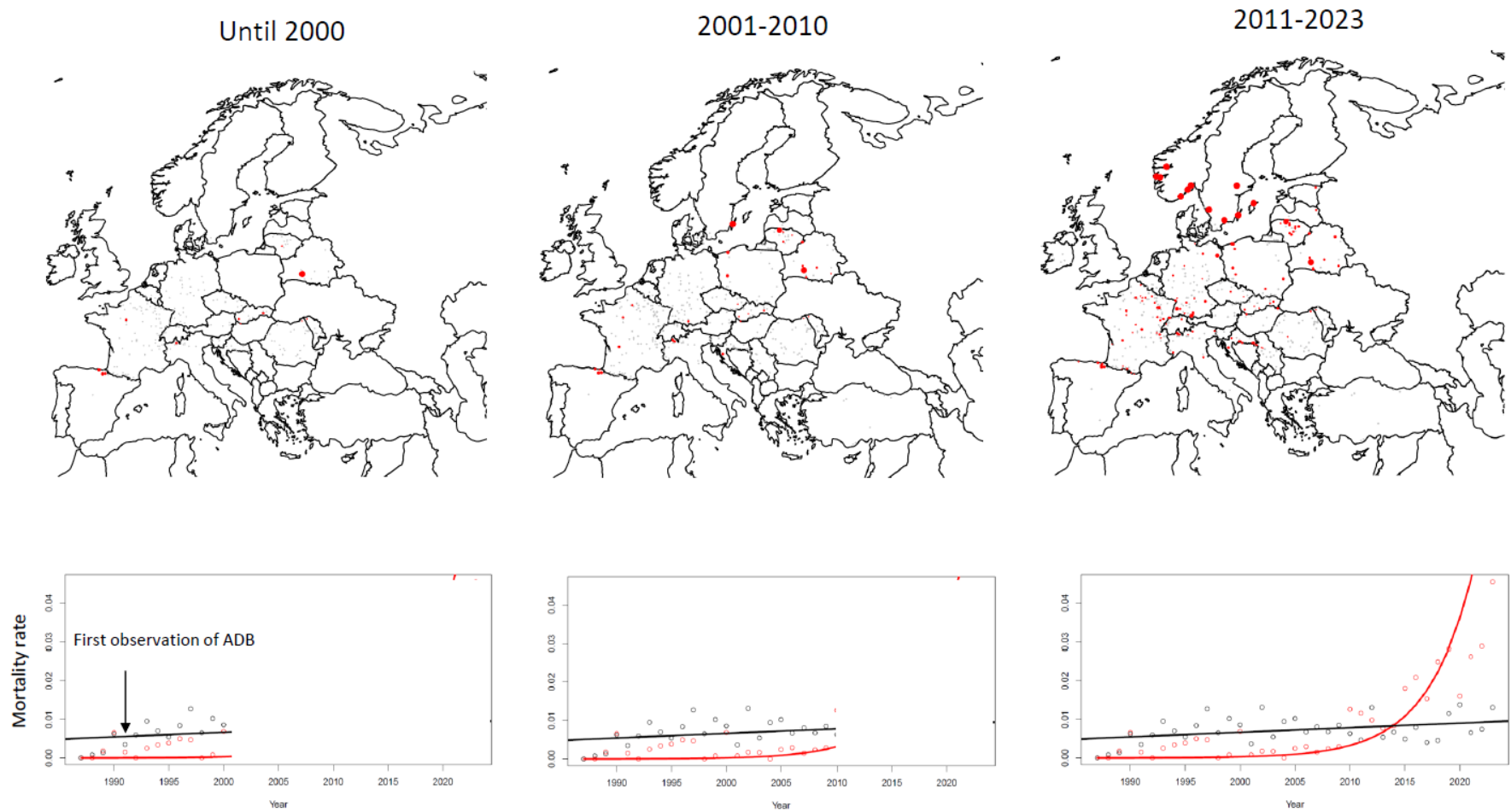

**Fig. 5:** Data from long-term forest health monitoring such as the ICP Forests level I dataset is able to monitor population decline at European level. The maps show how mortality in European ash and narrow-leafed ash has advanced from 1987-2023 in Europe. Size of the circles is proportional to mortality

rates. The same data is shown as points as European-wide average together with background mortality as a baseline. The graphics illustrate that despite the fact that ADB has arrived already in the beginning of the 1990s, it took considerable time until population decline has become visible at European scale. Ash mortality points (red) were fitted with an exponential function that explained approx. 40% of the temporal variation in mortality. A total of 407 ICP Forests Level I plots across Europe containing *Fraxinus excelsior* and *Fraxinus angustifolia* were analyzed for producing this graph. Note that despite the high mortality in Norway and Sweden only a very low number of ash trees were present in these plots between 2011 and 2023.

**Supplement:**

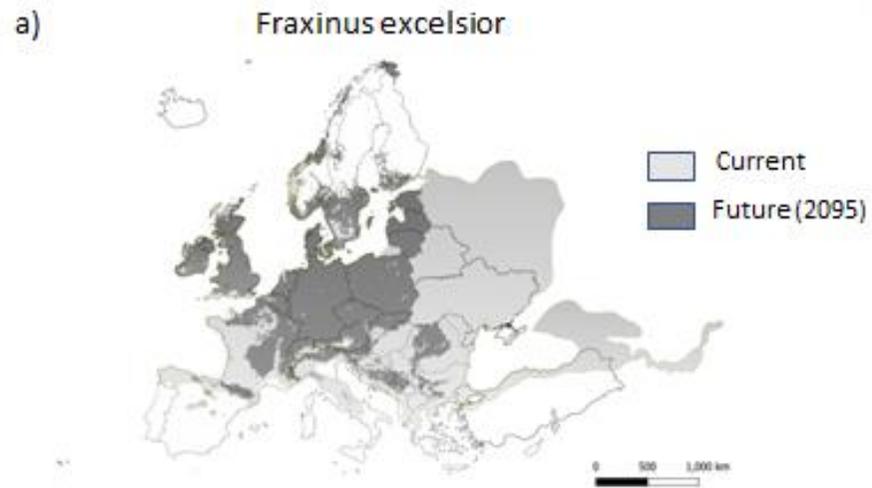
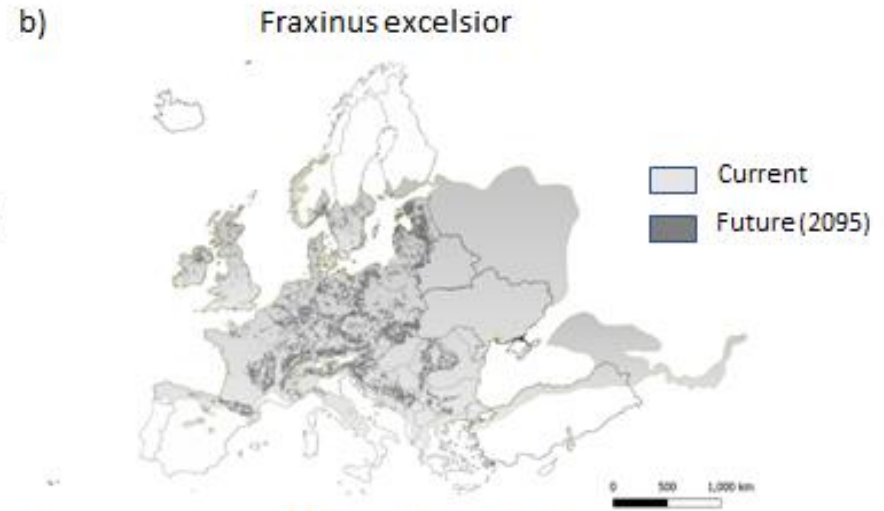
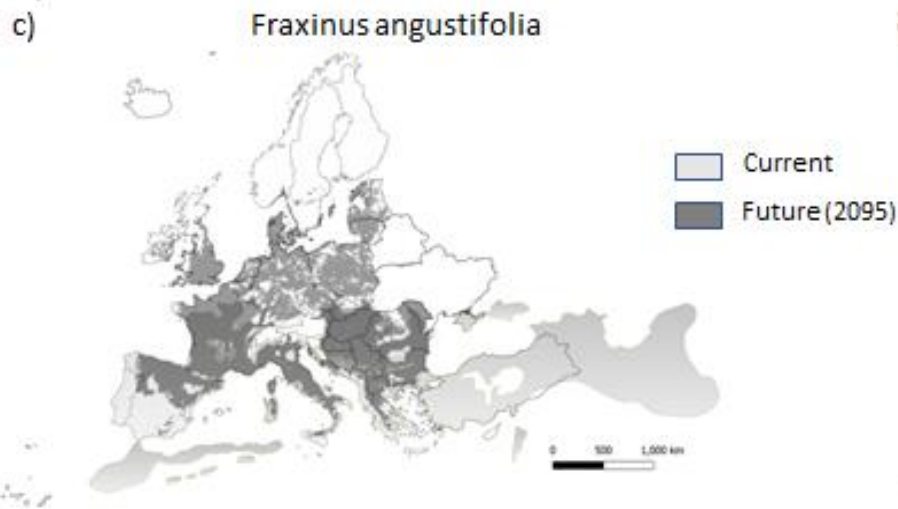
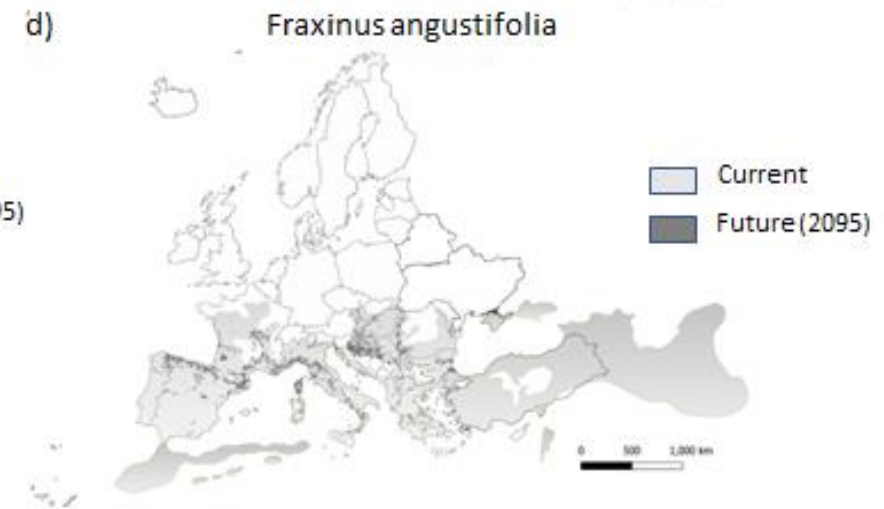

**Fig. S1:** Future projections for *Fraxinus excelsior* and *Fraxinus angustifolia* occurrence until 2095 (dark grey) under the RCP 8.5 (based on the data from Mauri et al. 2022). Subplots a) & c) show the projected occurrence when only climate variables are considered in the species distribution model (7 long-term bioclimatic variables from the EUROCORDEX regional climate simulations). Subplots b) and d) show model results for which land use and dispersal capacity were additionally considered besides climate variables. While decolonization at the southern range margin is likely to occur under both scenarios for both species, range expansion towards more northerly locations due to global warming seems to be highly unlikely under realistic dispersal scenarios. Note that these projections solely consider climate, while neither ash dieback nor emerald ash borer are currently accounted for in the models. Maps are projected to a 10km grid and represent binary probabilities of occurrence. Briefly, the data in Mauri et al. 2022 is based on a climatic ensemble mean of 11 regional climate models (RCM) projected until the end of the century and considering two different future climate scenarios (optimistic and pessimistic). Dispersal capacity for both ash species were integrated by using mean dispersal distances according to Tamme et al. 2014 and Thomson et al. 2018. For further details see Mauri et al. 2022. Current occurrences (light grey) were taken from EUFORGEN species distribution maps (Caudullo et al. 2022)